\documentclass[conference]{IEEEtran}
\IEEEoverridecommandlockouts
\makeatletter
\def\ps@headings{%
\def\@oddhead{\mbox{}\scriptsize\rightmark \hfil \thepage}%
\def\@evenhead{\scriptsize\thepage \hfil \leftmark\mbox{}}%
\def\@oddfoot{}%
\def\@evenfoot{}}
\makeatother
\usepackage{subfigure}
\usepackage{colortbl}
\usepackage{bm}
\usepackage{graphicx}
\usepackage{caption}
\usepackage[switch]{lineno}

\usepackage{graphicx}  % Written by David Carlisle and Sebastian Rahtz
\usepackage{url}       % Written by Donald Arseneau

\usepackage{amsmath}   % From the American Mathematical Society
\usepackage{cite}

\usepackage{amsfonts,amssymb}

\usepackage{stfloats}

\usepackage{cases}
\usepackage{algorithm}
\usepackage{multirow}
\usepackage{algorithmic}
\usepackage{stfloats}
\usepackage{epstopdf}

\makeatletter

\newcommand{\Rmnum}[1]{\expandafter\@slowromancap\romannumeral #1@}
\makeatother

\newcommand{\ls}[1]
    {\dimen0=\fontdimen6\the\font
     \lineskip=#1\dimen0
     \advance\lineskip.5\fontdimen5\the\font
     \advance\lineskip-\dimen0
     \lineskiplimit=.9\lineskip
     \baselineskip=\lineskip
     \advance\baselineskip\dimen0
     \normallineskip\lineskip
     \normallineskiplimit\lineskiplimit
     \normalbaselineskip\baselineskip
     \ignorespaces
    }

\usepackage[absolute,showboxes]{textpos}

\TPMargin{3pt}

\begin{document}
%\linenumbers

\newcommand{\copyrightstatement}{
    \begin{textblock}{0.84}(0.08,0.93) % tweak here: {box width}(left position, bottom position)
         \noindent
         \footnotesize
         \copyright 2022 IEEE. Personal use of this material is permitted. Permission from IEEE must be obtained for all other uses, in any current or future media, including reprinting/republishing this material for advertising or promotional purposes, creating new collective works, for resale or redistribution to servers or lists, or reuse of any copyrighted component of this work in other works. DOI: 10.1109/MNET.001.2100691.
    \end{textblock}
}
\copyrightstatement

\title{{\huge The Rise of UAV Fleet Technologies for Emergency Wireless Communications in Harsh Environments}}
\author{Zhuohui Yao,~\IEEEmembership{Student Member,~IEEE,} Wenchi Cheng,~\IEEEmembership{Senior Member,~IEEE,} \\Wei Zhang,~\IEEEmembership{Fellow, IEEE,} Tao Zhang, and Hailin Zhang,~\IEEEmembership{Member, IEEE}\vspace{0.2cm}
\\
\thanks{\ls{.5}
\setlength{\parindent}{2em}

This work was supported in part by  National Key R\&D Program of China under Grants 2021YFC3002102 and 2020YFA0711400, and the key R\&D Plan of Shaanxi Province under Grant 2022ZDLGY05-09.

Z. Yao, W. Cheng, and H. Zhang are with the State Key Laboratory of Integrated Services Networks, Xidian University, Xi'an, 710071, China (e-mails: zhhyao@stu.xidian.edu.cn; wccheng@xidian.edu.cn; hlzhang@xidian.edu.cn).

W. Zhang is with School of Electrical Engineering and Telecommunications, The University of New South Wales, Sydney, NSW 2052, Australia (e-mail: w.zhang@unsw.edu.au).

T. Zhang is with Xidian University, Xi'an and the National Earthquake Response Support Service, Ministry of Emergency Management of the People's Republic of China, Beijing, China (e-mail: zhangtao@cea-igp.ac.cn).
}

\vspace{-55pt}
}
\date{\today}

\maketitle

\thispagestyle{empty}

\begin{abstract}
For unforeseen emergencies, such as natural disasters and pandemic events, it is highly demanded to cope with the explosive growth of mobile data traffic in extremely critical environments. An Unmanned aerial vehicle (UAV) fleet is an effective way to facilitate the Emergency wireless COmmunication NETwork (EcoNet). In this article, a MUlti-tier Heterogeneous UAV Network (MuHun), which is with different UAV fleets in different altitudes, is proposed to flexibly serve various emergencies. We refresh the key performance indicators of full coverage, network capacity, low latency, and energy efficiency in harsh environments. Then, we present the special challenges regarding shadowing-dominated complex channel model, energy supply limited short-endurance, various communication mechanisms coexistence, and communication island for underground users in UAV-based EcoNet, followed by the MuHun-based EcoNet architecture and its advantages. Furthermore, some potential solutions such as the new hybrid-channel adapted resource allocation, reconfigurable intelligent surface assisted UAV communications, competitive heterogenous-networks, and magnetic induction based air-to-ground/underground communications are discussed to effectively achieve full coverage, high capacity, high energy efficiency, and diverse qualities of services for EcoNets in harsh environments.
\end{abstract}

\begin{IEEEkeywords}
\ls{1.0}
\setlength{\parindent}{2em}
Emergency wireless communications network, UAV fleet, reconfigurable intelligent surface, magnetic induction.
\end{IEEEkeywords}

\section{Introduction}

Over the past few years, a large number of emergencies, such as Mexico earthquake in 2017, Australian Bush fires in 2019, frequent hurricanes (Katrina, Gustav, Irene, and Isaac) in America, and COVID-19, etc., made a significant global impact on social and economic orders. All emergencies can be generally classified into two kinds of categories. The first is natural disasters such as earthquakes and hurricanes, where the traditional communication infrastructure is seriously disrupted or overloaded. It is challenging to communicate with each other in disaster areas at a limited time due to the sudden growth of data traffic and possible secondary disasters. The second is hot events like the outbreak of new coronavirus pandemic worldwide, where the explosive growth of multi-medium traffic critically challenges well-preserved backbone network infrastructure. During the pandemic period, communication services such as remote education, telemedicine, and video meeting, rapidly increase. Also, new applications like contactless delivery services are emerging. While the ubiquitous connectivity and efficient communications, which highly support the rescue efficiency, are challenged by rigid facilities and harsh environments. Actually, no matter natural disasters or hot events, it is unexpected and instantaneous with a mess.  There are a large number of people trapped or covered on the site. Besides, various rescue institutions participated in the search and rescue process at the frontline. The damaged/overloaded infrastructure and limited resources cannot support effective information exchange and sometimes even basic communications. Considering the characteristics of all emergencies, a temporary network architecture that quickly responds to the urgent communication requirement needs to be investigated.

In recent years, unmanned aerial vehicle (UAV) has been widely used in disasters such as earthquakes, landslides, and floods~\cite{6030903}. It is usually placed as a relay or temporary base station to accommodate the communication needs due to the flexible and three-dimensional mobility in emergency rescuing scenarios~\cite{8068199}. For different aerial platforms, there are various UAVs with different functions such as long-endurance UAVs at the high altitude, vertical-work UAVs at the medium altitude, and cooperative rotary-wing UAVs at the low altitude. Therefore, MUlti-tier Heterogeneous UAV Network (MuHun), which includes heterogeneous UAV fleets in different altitudes, is a flexible network architecture to cope with various emergencies. MuHun inherently has the advantages to energize Emergency wireless COmmunication NETwork (EcoNet) that provides full coverage and effective communication solutions for comprehensive services~\cite{9748064}. The key performance indicators (KPIs) in harsh environments different from that in general-purpose networks are refreshed as follows.

{\bf Full coverage} --- For emergencies, it is highly expected for MuHun-based EcoNets to serve the dead zones, where existing communications infrastructures are destroyed or overloaded. However, this kind of full coverage~\cite{8624410} requirement imposes new challenges for MuHun-based EcoNets. First, how to provide continuous communication services for multiple trapped users irregularly distributed in emergency areas. Second, in some disasters such as earthquakes, hurricanes, and mudslides, the trapped users are very likely to be masked by the obstacles/shadows or buried under the ground and very hard to obtain timely and stable connections. Therefore, it is necessary to exploit advanced techniques for achieving full coverage under MuHun-based EcoNets .

{\bf Network capacity} --- In disasters or hot events impacted areas, the instantaneous high-volume data traffic such as the dispatch instructions from the command center and the feedback information from the trapped users and the frontline rescuers, results in the overloaded communication infrastructure and network congestion. Also, the path-loss, fading, rain attenuation, and shadowing parameters seriously decrease the long-distance transmission rate due to the harsh environment in the accident areas. Therefore, the network capacity is a key indicator for the explosive growth of data traffic transmission. How to increase the network capacity with limited emergency communication resources is the second bottleneck.

{\bf Low latency} --- UAV ad-hoc networks, wireless local area networks, cellular mobile networks, and wireless sensor networks in the MuHun-based EcoNets, provide a variety of transmission mechanisms, network access protocols, and user terminals for rapid rescue. Besides, seamless information exchange and effective cooperation among various entities are vital for rescuing in accident areas. However, the end-to-end network latency, from the frontline to the command center, is constrained by the high mobility of rescuer, congested/interrupted base station, the trajectory of  UAV,  the location of survivor, the data transmission between different interface, etc. Therefore, the coexistence of HETerogeneous NETwork (HetNet) with sharing resources is very challenging for low latency.

{\bf Energy efficiency} --- Furthermore, numerous communication nodes including sensing nodes, positioning nodes, and trapped nodes are irregularly and densely distributed for serving the large-disaster area or hotspot area. It is very important for rescuing equipment to be lightweight, which imposes the small-size and zero battery related challenges in MuHun-based EcoNets. Also, with the development of new emergency communication paradigms, emergency communication applications have evolved from narrowband services like voices into broadband services like multi-medium streams. However, the energy resources are extremely limited in emergency areas, especially in natural disaster areas. Therefore, one of critical problems is the energy efficiency of MuHun-based EcoNets.

As a large-scale heterogeneous network with a complex structure, high flexibility, and dynamic topology, MuHun faces a variety of special challenges for emergency communications. Driven by the requirements of KPIs, MuHun-based EcoNets need refined designs to overcome the challenges.

\section{Challenges for UAV-based emergency wireless communications}

In general, emergency wireless communication network is a specially customized network for agile communications under the constraints of limited resources, harsh environments, and damaged infrastructures to cope with the exploding data traffic with strict latency requirement. Take geological disasters for example, it is not predictive of certain time, location, and damage severity. The rescue community believes that there is a Golden 72 hours after the disasters, during which the survival rate of victims is extremely high~\cite{6184055}. Every time an extra piece of soil is dug to give the injured a chance to breathe and live. Thus, the precise locations of trapped users need to be provided. The first barrier for emergency communication is to rapidly deploy a new communication system that facilitates commanding and rescue within Golden 72 hours, which is the consensus of the rescue researchers. During this period, all communication methods are available, such as satellites, emergency communication vehicles, UAVs, soldier radios, mobile phones, etc. Among them, the UAV can be not only used as an aerial base station to support communications, but also equipped with various sensing devices such as thermal imagers, infrared scanners, GoPro, etc., to collect disaster information. Therefore, UAVs play an important role in the emergency communication network due to their strong plasticity. There are some challenges that still need to be overcome.

\subsection{Shadowing-Dominated Complex Channel Model }

In UAV-based EcoNets, due to the wind, rain, fog, poor angle of inclination, and obstacles resulted masking impact, the traditional air-ground line-of-sight (LOS) communications assumption is not practical. The obstacles between the UAVs and the trapped users, resulting the severe shadowing, play an important role in the long-distance transmission. Especially after some geological disasters, the destroyed topography, as well as collapsed buildings, degrades the signal power at the receiver due to the signal scattering.
Therefore, the shadowing-dominated complex channel model should be considered. For example, the links from UAVs/terrestrial base stations to the trapped users undergo $\mathcal{F}$ composite fading~\cite{7886273} because the channel characteristics vary along with the unpredictable secondary disaster resulted by topography complexly changes. Also, the rain attenuation impact the long-distance communications due to harsh environments in the disaster area. The traditional channel model cannot depict the hybrid-channel model. Therefore, an accurate and tractable channel model to capture the complicated fading and shadowing parameters for UAV-based emergency wireless communications need to be formulated. Furthermore, a universal channel model is the basic characteristics and the corresponding resource allocation is very challenging in the UAV-based EcoNets.

\subsection{Energy Supply Limited Short-Endurance }

For emergency wireless communications, UAVs are commonly used for supplementary coverage to resist harsh environments, which challenges the endurance of UAVs.  During hovering, UAVs consume much energy to fight against self-gravity and provide momentum to move forward. Since it is impossible for UAVs to always fly at a constant speed, the high current discharge is required when encountering strong winds or higher speed requirements. Moreover, UAV communication performance guaranteeing must be taken into consideration, thus causing indispensable energy consumption. Therefore, the battery life of UAVs is very related to the stability of EcoNets. Nowadays, multiple-input multiple-output (MIMO) assisted UAVs are usually considered to increase the network capacity of emergency wireless communications.  However, they cannot maintain long hours of work due to the limited energy supply resulted by the physical structural properties of UAVs. Under extremely harsh weather,  UAVs are usually equipped with blade protection covers that are suitable for strong wind and rain. The blade protection cover increases the weight and power consumption of UAVs. For exploiting the flexibility and long-endurance of UAVs, it is not recommendable to carry out overweight and power-consumed devices such as spare batteries and multiple antennas with large-size. Apparently, energy efficiency is very important for UAV networks, as most energy has to be used for flying for a long time running. To overcome this challenge, both reducing mechanical energy consumption and transmit as well as signal processing power need to be considered.  We proposed to solve these problems with two aspects.
     \begin{itemize}
    \item Different types of UAVs are adopted to match the different tasks. For example, the large-scale UAVs with long-endurance covers most of the disaster areas in high altitude. It is robust to resist the harsh weather with its large size. However, the number of this type UAV is limited since the costs can not be ignored due to the sophisticated hardware fabrication and fuels-motivated energy. Then, a few of low-cost UAVs are used to perform auxiliary tasks, such as monitoring, mapping, and positioning. Under such a multi-UAV scenario, effective cooperation can reduce energy consumption by optimizing trajectory of these UAVs, which is specified in Section~III.
    \item Reduce the load of medium/small UAVs which are used to perform the short-time mission in low altitudes. For example, replacing the large-scale antenna with a cable or some light reflective surface. The UAV with a cable can supply constant energy from the ground power bank, thus supporting remarkable motivation to perform the time-consuming tasks in the harsh environments. As for the UAVs with reflective surfaces, a major concern is how to achieve reliable communications with energy harvesting simultaneously, which is specified in Section~IV.
    \end{itemize}

\subsection{Various Communication Mechanisms Coexistence}

During the command and rescue process, communication latency is a very critical performance indicator. Since most emergencies occur in a closed area with a short process, emergency communication systems need to be self-contained. It is required not only to keep in touch with the outside, but also to ensure timely communication in post-disaster areas. However, with the addition of new emergency communications paradigms, the types of services are expanded and the network congestions are frequently triggered due to mixed communication mechanisms and concentrated data traffic in MuHun-based EcoNets. As a multipurpose system, there are various networks in EcoNets, among which all nodes share the limited resources to connect with each other excepting specially licensed bands for the private network. In addition, a piece of information exchange between different network is strongly dependent on the network infrastructure, which drives the convergence in the network layer. Therefore, the resource management for interconnection of HetNets is very challenging.

\subsection{Communication Island for Underground Users}

When natural disasters such as earthquakes or accidents such as mine collapses occur, some users are trapped underground. The communication island is caused because the underground users are difficult to obtain rescue information. Under such scenarios, UAVs carried with other payloads are used to search and detect vital signs. Critically, the trapped nodes need to communicate with or at least be sensed by the UAVs, which are inspecting the disaster or common underground areas. The underground environment, which contains soil, rock, and water, is very challenging for wireless communications. For such a mixed-media communication environment, followed by severe path-loss, it is very hostile to communicate with underground users. Clearly, the well-established wireless propagation techniques for terrestrial networks do not work well~\cite{5452976}. Thus, a solid technology is urgently required to guarantee that signals can penetrate into the ground for Air-to-Ground/UnderGround (A2G/A2UG) communications.

\begin{figure*}
\centering
\includegraphics[scale=0.31]{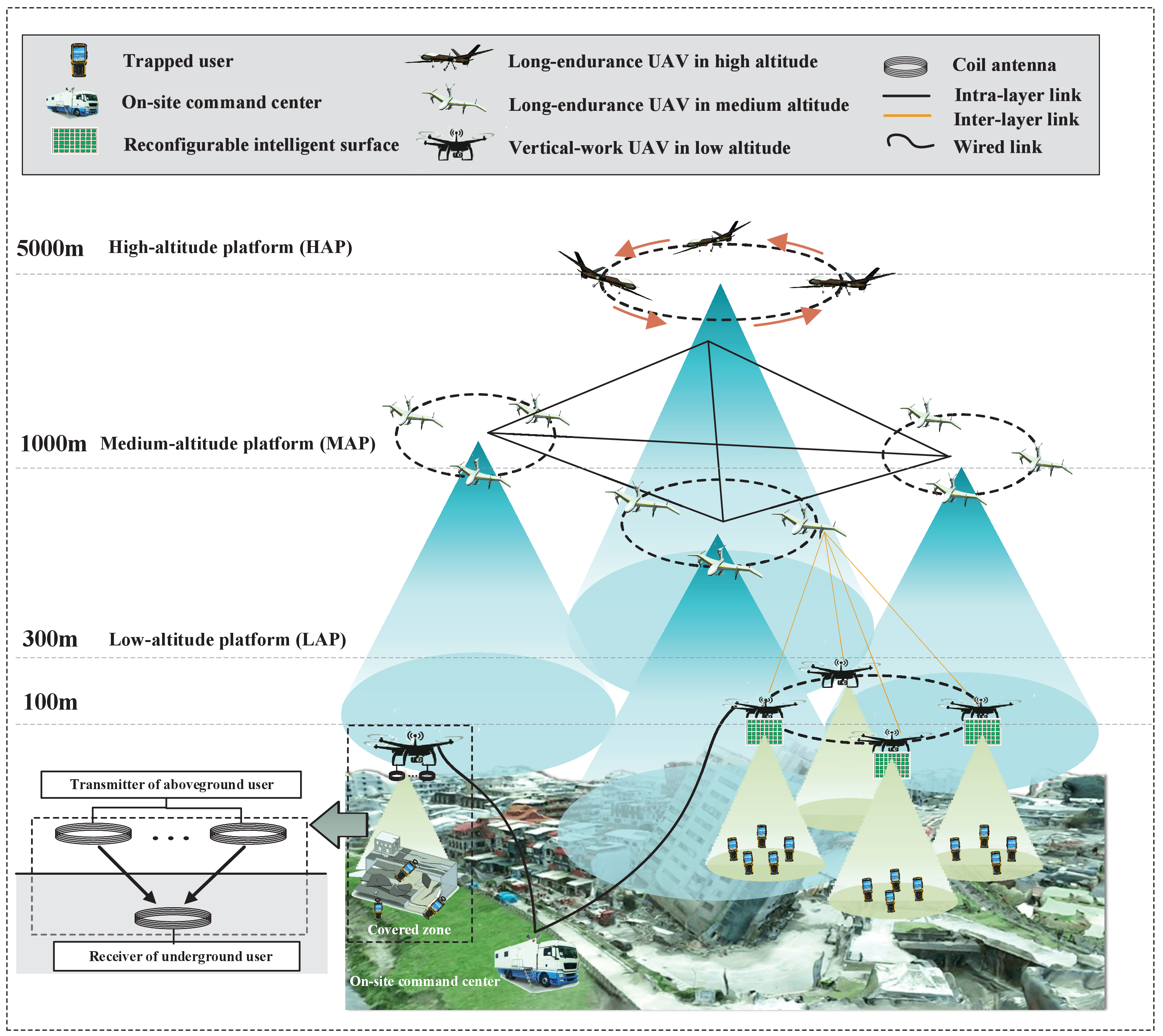}
\caption{The MuHun-based emergency wireless communications network architecture.} \label{fig:ECNA_model}
%\vspace{-0.2cm}
\end{figure*}

\section{MuHun-based EcoNet Architecture and its advantage}

We propose the MuHun-based EcoNet architecture for post-disaster rescue, as shown in Fig.~\ref{fig:ECNA_model}, which consists of long-endurance UAVs in high-altitude/medium-altitude, reconfigurable intelligent surface assisted UAVs (RIS-UAVs)/coil antenna equipped UAVs (coil-UAVs) in low altitude, on-site command center, and trapped users with irregular distribution. Under such an emergency scenario, where the infrastructures are destroyed, heterogeneous UAV fleets in different altitudes can execute various tasks. Long-range and long-endurance UAVs above $1000$ meters altitude have the advantages of rapid arrival and stable hovering in the air. The core links among UAVs, shown as intra-layer links in Fig.~\ref{fig:ECNA_model}, are established since long-endurance UAVs cover most of sparse area and their strong survivability support network robustness. For medium and short-range UAVs below $300$ meters altitude, which take-off and land vertically without a runway, they are convenient to plan reconnaissance and ground operations. These nodes alternately access the UAV ad-hoc network, shown as inter-layer links in Fig.~\ref{fig:ECNA_model}. On the other hand, the UAVs with tethering cables, called tethered UAVs~\cite{7158825}, are motivated by the power bank on the ground. They are widely utilized in stable hovering tasks such as precision mapping,  agricultural monitoring, and disaster response missions without being limited to the on-board electric storage cells (batteries). Tethered UAVs can work at several-hundred meters of altitude with hovering in the air for a long time, thus facilitating terrestrial communications for the difficult rescue.  The location of on-site command center is reasonably selected to facilitate communications for trapped users. Some of trapped users on the ground are connected to the UAVs/terrestrial base stations. The other trapped users are located in covered zones, where they are masked or buried under the ground. RIS-UAVs can improve the communication environment for the trapped user with high energy efficiency. Magnetic induction (MI) based transmission makes the coil-UAV enable connecting with the underground users, thus enhancing A2G/A2UG communications. Combining the long-endurance UAVs, vertical-work UAVs, RIS-UAVs, and coil-UAVs, heterogeneous UAV fleets for post-disaster rescue are established.

Under emergency scenarios, the burst service flows generated by various equipment are required to communicate at a strict delay requirement by MuHun where UAVs handoff frequently happens based on the instantaneously changing channel condition, task allocation, and operation. Under such a hybrid UAV situation with multiple functionalities, how the different types of UAVs communicate and cooperate is worthy of attention.

For achieving the communication and cooperation among UAVs, the first barrier is the deployment optimization of UAVs, which includes placement and trajectory optimization. It aims to minimize the number of required UAVs and the movement distance of UAVs to provide wireless coverage for irregularly distributed ground/underground terminals in the MunHun-based EcoNet without fixed terrestrial infrastructure. During the deployment optimization, the features of different UAVs, user density, and target tasks are considered to ensure that each ground/underground terminal is within the communication range of at least one UAV.

Next, the second barrier is how to establish the communication links for UAVs. At present, the coordination among different types of UAVs are usually hard and it is nearly impossible to establish a ``face-to-face'' relationship because there are several levels between the frontline UAV and command agency. This difficulty delays the information exchange and cooperation, which hinders dealing with the air task in a fast and efficient manner. Therefore, a flat communication mechanism for MunHun-based EcoNet is necessary, which encourages the UAVs equipped with the unified communication module. Then, the UAV on a mission can acquire required sensor images and data directly from other UAVs through the communication link, which also helps the joint trajectory optimization among UAVs to achieve more expanded wireless coverage. It is worth mentioning that most of spectrum resources are available in the post-disaster rescue process, thus likely causing interference and waste. For different altitudes of UAVs, the propagation links undergo diverse fading and shadowing. Therefore, the spectrum allocation based on corresponding channel conditions needs to be discussed. High frequencies are distributed to the LOS links and low frequencies are distributed to the terrestrial and shadowed links. Furthermore, multi-channels in the same links also need to allocate the spectrum reasonably which corresponds to the channel condition and power allocation, as shown in Section~IV-A.

As for the synergy effect, UAV classification is driven by different functionalities, such as computation, cache, and control, thus facilitating the collaboration of position, sensing, and communications. After determining core nodes with licensed controlling, partially distribution with introducing mobile edge computation and storage units nearby is considered to accelerate the major download and service mission. By deploying the mobile edge computation and storage resource around the control center, it makes full use of the location, channel status, and terminal information provided by the wireless access network to realize the positioning, fast calculation, and feedback services adapted to communication environments. To further enhance the performance of UAVs' communications and cooperation, jointly optimizing trajectory, beamforming, and power control~\cite{8068199}~\cite{9351782} based on the UAVs' positions, phase-shift of antenna, and resource allocation is an effective solution. Actually, constrained by the complete coverage, required transmission rate, and peak transmit power, alternative optimization schemes, as well as artificial intelligence method, are common to be used. Additionally, due to the flexibility of nodes, MunHun-based EcoNet may fail to work since the links are frequently switched and UAVs may be damaged by the harsh environment. An adaptive routing plan is needed to stabilize the link quality of core nodes and update interrupted nodes.

\section{Further solutions for MuHun-based EcoNets}

For such a large amount of  UAVs in MuHun-based EcoNets special challenges such as complex channel model, short-endurance, various communication mechanisms coexistence, and communication island for underground users, should be overcome under the conditions of harsh environment, limited resources, the explosive growth of traffic, and diverse services. It is urgent to re-exploit and integrate advanced beyond 5G and future 6G techniques, including new hybrid-channel adapted resource allocation, RIS-assisted UAV communications, competitive HetNets, and MI-based A2G/A2UG communications.

\subsection{New Hybrid-Channel Adapted Resource Allocation}

\begin{figure*}
\centering
\includegraphics[scale=0.6]{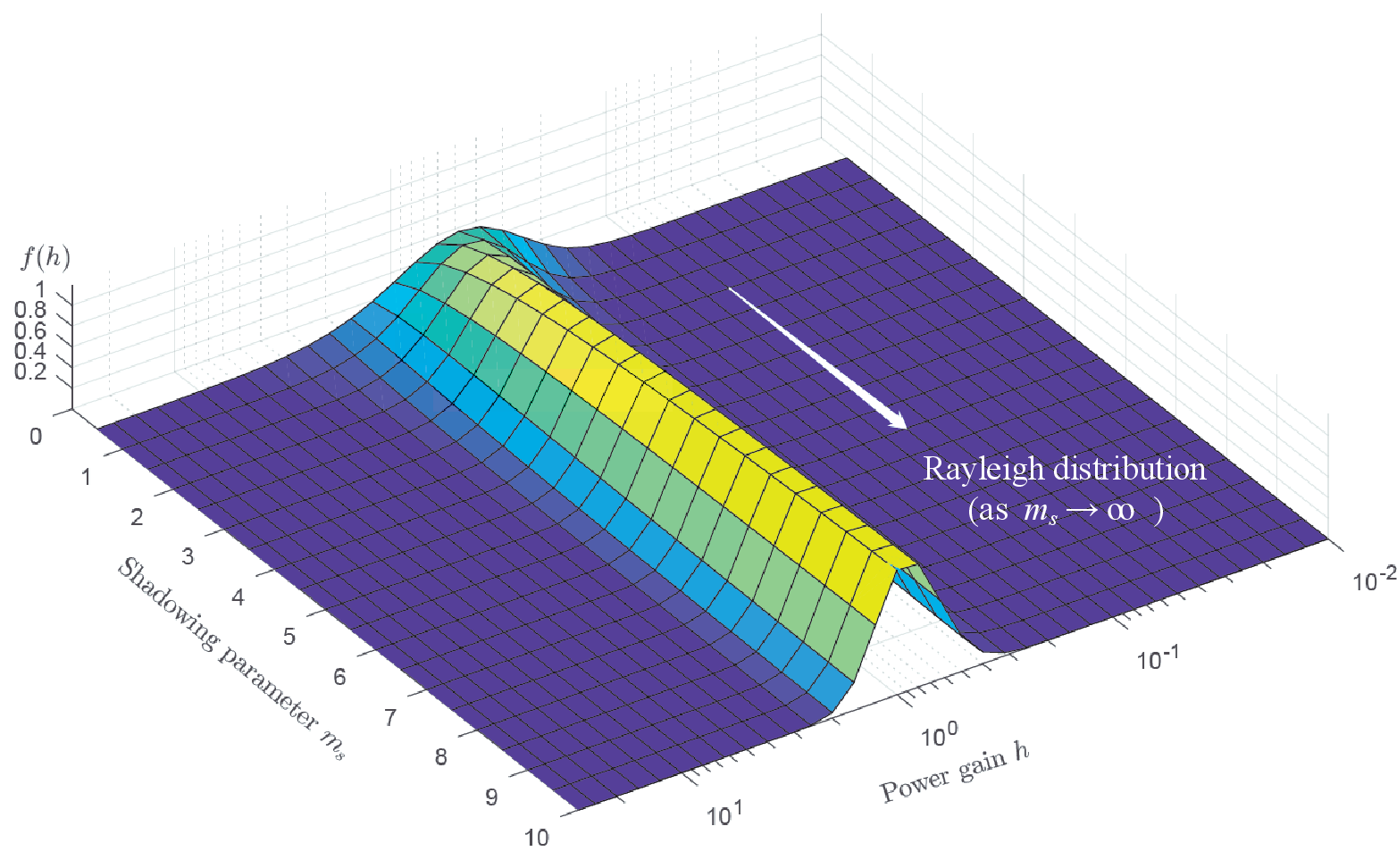}
\caption{PDF of shadowing-dominated channel.} \label{fig:Channel}
\vspace{-0.2cm}
\end{figure*}

\begin{figure*}
\centering
\includegraphics[scale=0.8]{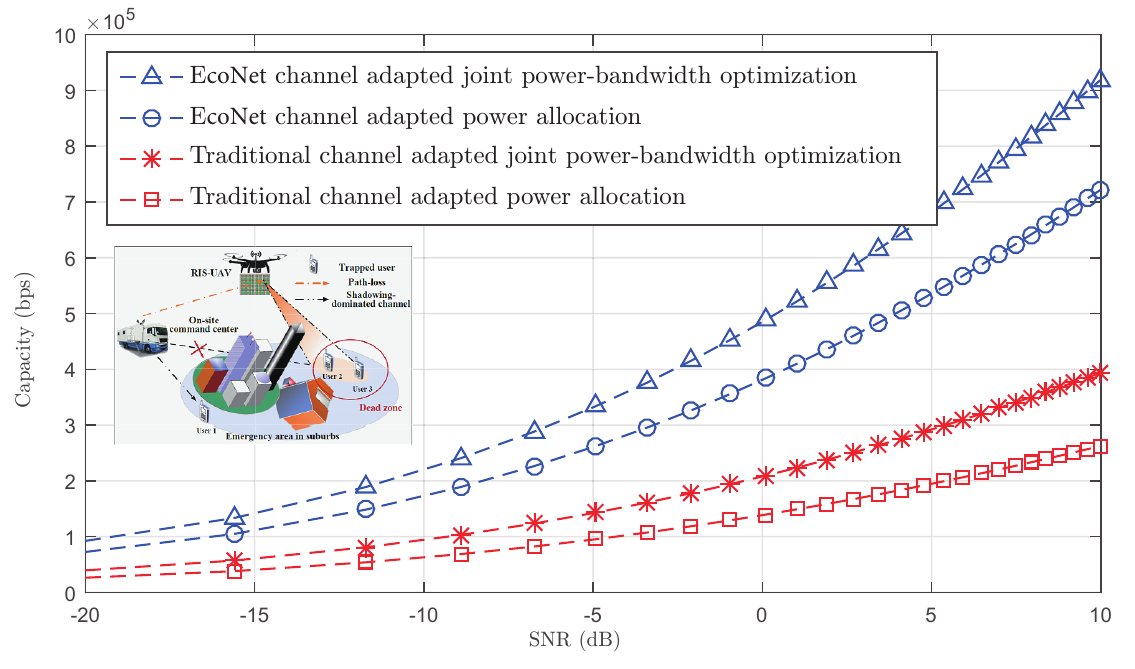}
\caption{The capacities with the EcoNet channel adapted resource allocation schemes (including joint power-bandwidth optimization and power allocation) and traditional channel adapted resource allocation schemes (including joint power-bandwidth optimization and power allocation).} \label{fig:RIS-UAV}
\vspace{-0.2cm}
\end{figure*}

In view of such complex channel conditions, it is very critical to character the shadowing parameter $m_{s}$ and derive the new hybrid-channel model. The shadowing parameter $m_{s}$, which follows inverse Nakagami-$m$ distribution in  $\mathcal{F}$ composite fading model,  represents severe shadowing as $m_{s}\rightarrow0$ and weak shadowing as $m_{s}\rightarrow\infty$. To depict the difference between new hybrid-channel and traditional channel, Fig.~\ref{fig:Channel} shows Probability Density Function (PDF) of shadow-dominated channel with key factors, i.e. shadowing parameter and power gain. As shown in Fig.~\ref{fig:Channel}, the probability of large power gain increases as $m_{s}$ gets large values, which means the channel condition gets better as $m_{s}\rightarrow\infty$ corresponding to Rayleigh distribution. Based on the complex channel model, the adaptive resource allocation schemes are proposed~\cite{9453842}. Fig.~\ref{fig:RIS-UAV} compares the capacities of EcoNet channel adapted resource allocation schemes and the capacities of traditional channel adapted resource allocation schemes. It shows the application scenario of RIS-UAV communications over MuHun-based EcoNets. The link between on-site command center and trapped users is partly interrupted since the collapsed buildings cause a blind spot for communication, shown as dead zone. In this scenario, RIS-UAV is deployed between the on-site command center and trapped users, to passively reflect the transmit signal with the optimal phase shift. The limited power and bandwidth are allocated corresponding to changing EcoNet channel.  In the power-optimization simulation part, the EcoNet channel is assumed with the path-loss exponent of $0.3$ for the link of on-site command center and RIS-UAV, the fading parameter of $1$ and the shadowing parameter of $0.5$  for the link of RIS-UAV and the trapped users. Furthermore, the joint power-bandwidth optimization is performed with the alternating optimization approach, where the step factor is assumed as $0.01$. It illustrates that the EcoNet channel adapted joint power-bandwidth optimization scheme can achieve higher capacity than traditional schemes.

\begin{figure*}
\centering
\includegraphics[scale=0.72]{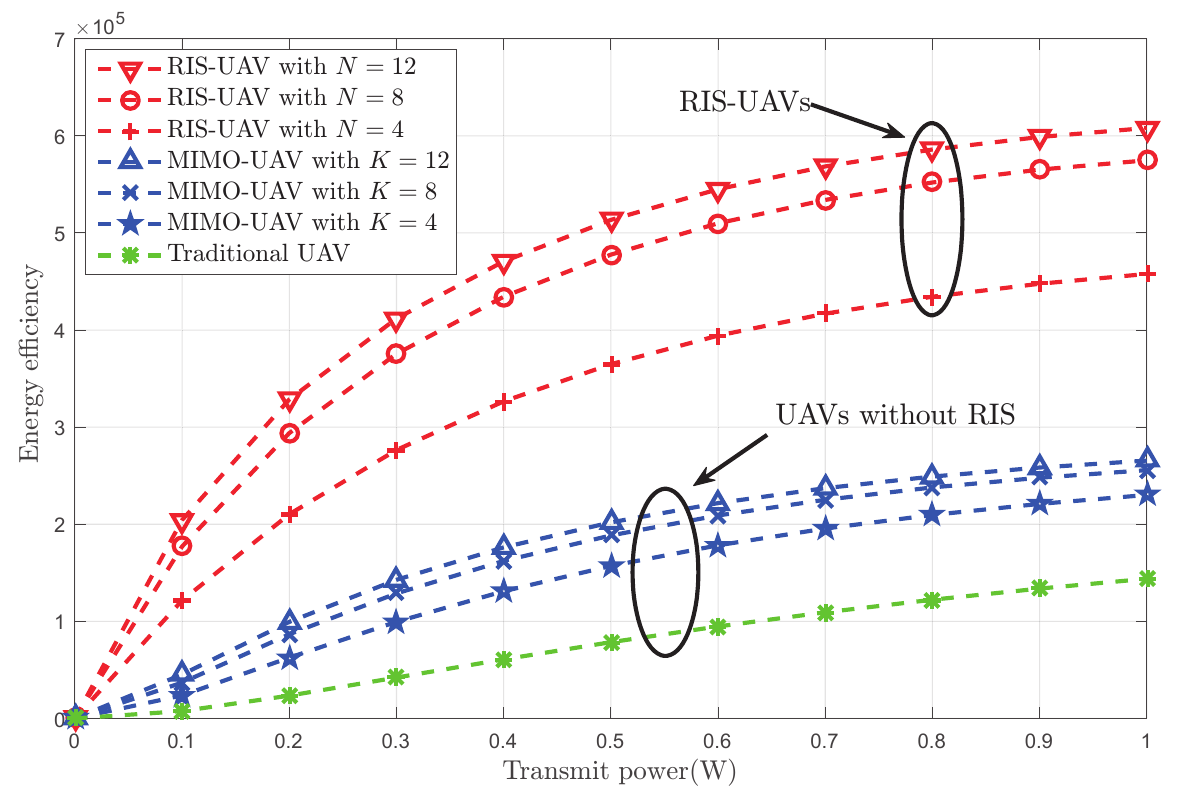}
\caption{The energy efficiencies with different numbers of traditional UAVs and RIS-UAVs.} \label{fig:EE}
\vspace{-0.2cm}
\end{figure*}

\subsection{RIS-Assisted UAV Communications}

From energy efficiency perspective, since RIS can easily be deployed and smartly reconfigures the wireless propagation environments with very low energy consumption, Aerial RIS (ARIS) was proposed~\cite{9351782} to extend coverage with panoramic/full-angle reflection, in which RIS is integrated with UAVs/balloons (Here, RISs are not limited to being hung onto the UAV, they can also be equipped onto the UAV facade, which makes RIS and UAV integrated).  It is worth mentioning that each small scatter can be programmed by digital sequences to achieve real-time control of electromagnetic parameters without extra power consumption~\cite{8910627}. Following that, the lightweight integration and passive beamforming of RIS are helpful for UAV communications. Moreover, RIS-assisted communications services and applications have been investigated under different communication systems~\cite{9174910}, which verify the potential of capacity and energy efficiency enhancement. Therefore, the RIS-UAV communication is suitable to refine the performance of MuHun-based EcoNets. The performances of energy efficiency with MIMO-UAVs and RIS-UAVs are evaluated in Fig.~\ref{fig:EE}, where $K$ denotes the antenna number of MIMO-UAVs, $N$ denotes the reflective element number of RIS-UAVs. The power consumption of MIMO-UAV includes the circuit blocks, FPGA on the relay, and the power amplifier, which values in watts. While the power consumption of RIS-UAV comes from the FPGA on RIS and the PIN diodes, which values in milliwatts. Therefore, the total power consumptions with RIS-UAV cases are very low since there are no radio-frequency (RF) chains which are replaced by the PIN diodes with low power consumptions. As shown in Fig.~\ref{fig:EE}, it is clear that the RIS-UAVs can achieve higher energy efficiency than traditional UAVs. Although RIS-UAV is very promising to enhance the network performance and is expected to replace the MIMO-UAV under such scenarios, it still faces new challenges. First, the hardware design for RIS-integrated UAV needs to consider the compatibility of stability, reliability, and flexibility. Then, the joint-design of UAV altitude, trajectory, and passive beamforming related to the users' distribution for ARIS is more complex than that for terrestrial RIS. In addition, although ARIS prevents the signal from experiencing severe attenuation via smart reconfiguration, the dynamic channel estimation still makes a critical bottleneck due to the high mobility of ARIS. The above challenges deserve further investigation. On the other hand, a crucial issue of RIS-UAV is the computation of reflection coefficients (or beamforming matrix), which consumes UAV's battery. To solve this problem, we propose to further increase the energy efficiency of RIS-UAVs from the following two aspects: energy supplementary and computation task offloading. For energy supplementary, it is a simple and direct way to be assisted with the tethered UAV, which is motivated by the power bank on the ground via a tethering cable but limited with the given flight space. RIS mounted on the tethered UAV can enhance the coverage performance mutually thus solving the increased energy consumption of RIS's computation as well as helping flight trajectory of tethered UAVs. Besides, energy harvesting from solar and wind is an available solution in emergency areas, which generates energy by making full use of ambient environments. However, it is limited by the randomly changing weather, thus appearing RF based energy harvesting strategies. To enhance the RIS-UAV endurance, simultaneous wireless information and power transfer method can be considered. Since the reflective elements of RIS can be designed as separate controllers, they are divided into the energy-collect part and signal-reflect part. Hence, the former can be used to collect energy from the received RF signals and the latter can be used to reflect signals, which facilities the energy efficiency of RIS-UAV. For computation task offloading, mobile edge computing (MEC) is a well-known method to enable users to offload their computation-intensive and latency-sensitive tasks to edge servers, which can be played by other cooperative UAVs or remote access points. Benefits from the collaboration of UAV fleets, another UAV can act as a MEC server to help the task bits of the users for computing, thus saving the computation resource of RIS-UAV.

\begin{figure*}
\centering
\includegraphics[scale=0.42]{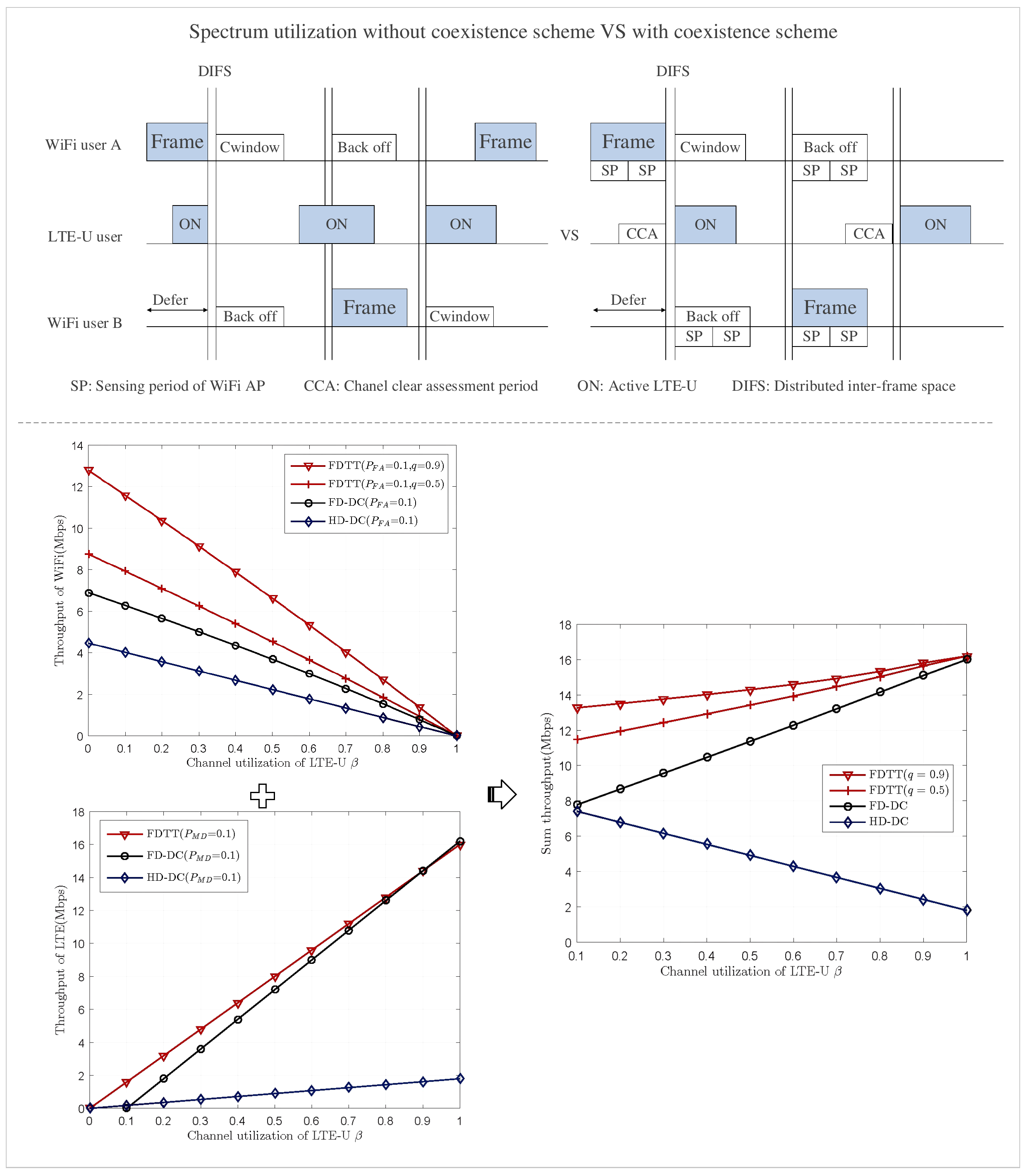}
\caption{Spectrum utilization for competitive HetNet and throughput performance evaluation for the proposed coexistence scheme. (The numerical results shows the individual throughput and sum throughput performance with channel utilization of LTE-U, where FDTT denotes Full-duplex based two-threshold methods, FD-DC denotes Full-duplex based duty-cycled methods. HD-DC denotes Half-duplex based duty-cycled methods, $P_{FA}$ denotes false alarm probability, $P_{MD}$ denotes miss detection probability, and $q$ denotes the probability that detected energy falls between two thresholds.)} \label{fig:Co}
\vspace{-0.2cm}
\end{figure*}

\subsection{Competitive HetNets}

Thank to the potential benefits of UAVs in enhanced performance, wide application, and cost-effectiveness,  they perform diverse missions with different payloads in the MuHun-based EcoNet. Under such scenarios, UAVs play the roles as base-stations, access points, relays, and users, which demands different but stringent communication KPIs. For UAV-assisted wireless communications (i.e., UAVs play as base-stations, access points, and relays), WiFi frequency bands are usually exploited to restore UAV-ground communications except utilizing the regulated frequency band to release specific commands. However, cellular users are facing limited transmission rates due to the limited spectrum and growing data traffic. Nowadays, more and more institutions are pursuing to offload the increasing cellular traffic to the unlicensed bands. Thus, the coexistence of WiFi and LTE in unlicensed bands (LTE-U) has been investigated.  The dynamic spectrum sharing scheme based on the load detection is one of the effective solutions. As shown in Fig.~\ref{fig:Co}, the full-duplex based two-threshold spectrum sensing method and further MAC protocol increase the spectrum utilization by avoiding frequently collision and spectrum wasting, thus increasing the throughput of HetNets as well as guaranteeing the individual performance of LTE-U and WiFi. To further enhance the co-existence performance in harsh environments where the bands and link qualities are unpredictable changing with the damaged infrastructure, deep reinforcement learning over mobile edge nodes can be used to optimize the channel resource allocation. With this scheme, the environment state can be modeled with time-varying channel information, active LTE-U users and WiFi users. Then, perform the environment state as inputs to the action-selection model. The actions are followed to allocate the channel resources for the active users, whose results along with unprocessed service are feedback to the inputs. Finally, a reward, the throughput and scheduling requirements of network, is used to guide a good learning loop of channel allocation strategy. Therefore, it is convenient to obtain real-time resource allocation strategies by learning from past experience of interacting with environment, thus increasing the throughput of HetNets. Whereas, for the future application to the mission-critical emergencies, the learning durations and service qualities are required to be elaborative, thus deserving to collaborate with other sensing and positioning information to complete the environment states. On the other hand, for cellular-connected UAV communications (i.e., UAVs play as users)~\cite{8470897}, the emerging spectrum interference is a critical problem. The spectrum scheduling reworks to the regulated bands and serves the high-priority missions with given security and sensitive scenarios. In addition, the diverse functions with corresponding payloads are required to support by high data rate, which also needs to share the cellular spectrum. While RIS-assisted UAV can adaptively design the three-dimension beamforming with the given channel state information. Furthermore, the integrated sensing and communications based no-orthogonal multiple access method can also be a promising solution.

\begin{figure}
\centering
\includegraphics[scale=0.6]{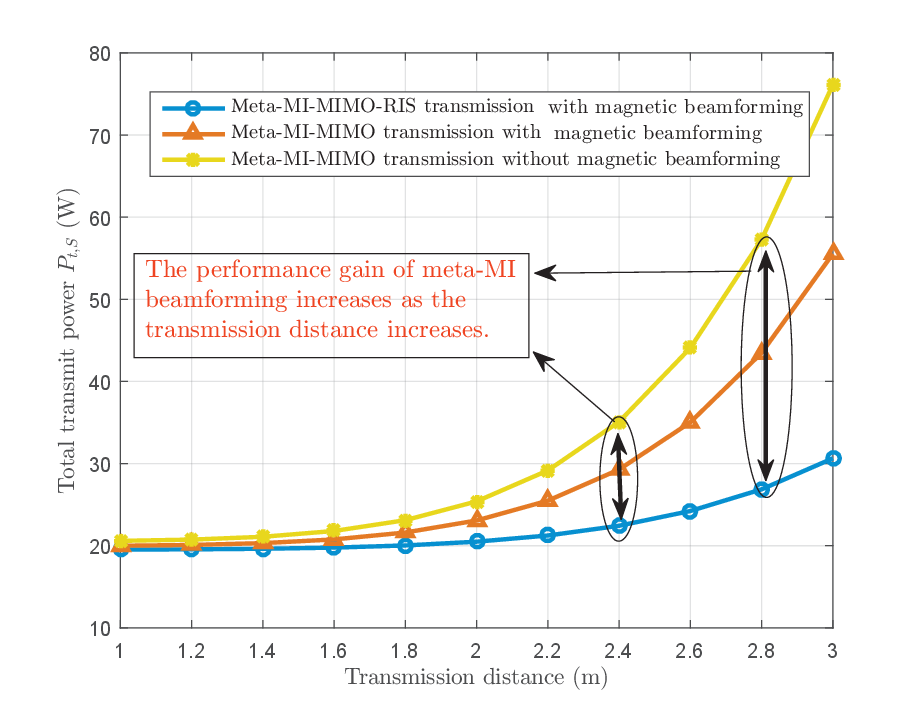}
\caption{Performance evaluation for meta-MI-MIMO-RIS transmission with magnetic beamforming, meta-MI-MIMO transmission with magnetic beamforming, and meta-MI-MIMO transmission without magnetic beamforming.} \label{MI_BF}
\vspace{-0.2cm}
\end{figure}

\subsection{MI-Based A2G/A2UG Communications}

To guarantee the efficiency for A2G/A2UG communications, metamaterials enhanced MI (meta-MI) is a feasible and effective solution~\cite{7272068}~\cite{MI_RIS}. Following Faraday's electromagnetic induction law, MI based transmission employs coil antennas to deliver power and information. As shown in Fig.~1, modulated sinusoidal currents in the transmit coils first induce the time-varying magnetic field in the space, by which sinusoidal currents are then correspondingly induced in the receive coils. In fact, the MI-based A2UG communication is a kind of near field communication (NFC), which is equipped with most of handhold-devices, such as mobile phone. Consequently, the information can be retrieved by demodulating the induced currents. Coil units with active circuits can be used to formulate the meta-MI reconfigurable surface, which can perform magnetic beamforming and concentrate the induced magnetic fields at underground users by adjusting the parameters of active circuits. The specific channel properties in MI-based A2UG scenario can be well estimated with the known coil size, distance, angle, and almost close permeability~\cite{7272068}. As shown in Fig.~\ref{MI_BF}, the performance gain of meta-MI beamforming increases as the transmission distance increases. Also, the magnetic fields around transceivers can be enhanced by the metamaterial shell which encloses the coil antenna. To extend the communication distance, multiple meta-MI coils without energy sources can be used to formulate a waveguide network which assists the data delivery from UAV to underground. Due to the limited bandwidth of coils, the transmission rate of MI communication is relatively low as compared with conventional electromagnetic (EM) wave based communications. The rate enhancement for A2G/A2UG communications can be obtained using meta-MI MIMO coil array, meta-MI magnetic beamforming, meta-MI orthogonal-frequency-division-multiplexing (meta-MI-OFDM), and meta-MI full duplex (meta-MI-FD). Also, the adaptive coding, modulation, and resource allocation schemes can be also utilized to increase the transmission rate.

\section{Conclusions and Future Work}\label{sec:conc}

MuHun-based EcoNets play an important role in the field of emergency rescue. This article explored the potential solutions by leveraging the emerging beyond 5G and future 6G technologies to achieve full coverage, high capacity, low latency, and high energy efficiency for MuHun-based EcoNets. In particular, the challenges regarding shadowing-dominated  complex channel model, energy supply limited short-endurance,  various communication mechanisms coexistence, and communication island for underground users have been elaborated. To overcome the problems, we proposed new hybrid-channel adapted resource allocation, RIS-assisted UAV communications, competitive HetNets, and MI-based A2G/A2UG communications, thus supporting the refreshed KPIs. To further refine this MuHun-based EcoNets, more future work is emerging.

Most of research focuses on communication systems at-disaster areas. In the previous discussion, we are aiming at the channel model on-site, wireless coverage, spectrum efficiency, energy efficiency, and data rate. However, pre-disaster warning, risk assessment, and post-disaster construction are also important. Therefore, intelligent EcoNet, which means aggregating all the emergency resources for timely perception, processing, communication, and decision-making, is our ultimate goal. From disaster prediction and risk assessment to post-disaster reconstruction,  there are several tasks that we still have to do. As a critical link for signal propagation, the intelligence of channel modeling is an important part. Besides, there are a large number of sensors deployed in places where disasters frequently burst, which collect terrain and weather information in a timely manner, and perform data transmission and processing constantly. Furthermore, due to the participation of different communication mechanisms, HetNet integration impacts the quality of service. Therefore, the diverse services including underground services, high-priority services, and so on, are required to be covered. We hope that this article can arouse the interests of researchers to re-farm more novel technologies for solving major issues of EcoNets.

\bibliographystyle{IEEEbib}
\bibliography{References}

\end{document}